\documentclass[oldversion]{aaa}
\usepackage[switch]{lineno}
\usepackage{graphicx}
\usepackage{txfonts}
\usepackage{natbib}
\renewcommand{\tabcolsep}{1.4mm}
\bibpunct{(}{)}{;}{a}{}{,} 

\def\ecs{erg~cm$^{-2}$\,s$^{-1}$}
\def\lum{erg~s$^{-1}$}
\def\bron{1RXS J171824.2--402934}

\begin{document}

\title{Monitoring campaign of 1RXS J171824.2--402934, \\
the low-mass X-ray binary with the lowest mass accretion rate}

\titlerunning{Monitoring campaign of \bron}
\authorrunning{J.J.M. in 't Zand et al.}

\author{J.J.M.~in~'t~Zand\inst{1}, P.G. Jonker\inst{1,2}, C.G. Bassa\inst{1,3},
  C.B. Markwardt\inst{4,5} \& A.M. Levine\inst{6}}

\offprints{J.J.M. in 't Zand, email {\tt jeanz@sron.nl}}

\institute{     SRON Netherlands Institute for Space Research, Sorbonnelaan 2,
                3584 CA Utrecht, the Netherlands
	 \and
                Harvard-Smithsonian Center for Astrophysics, 60 Garden Street,
                Cambridge, MA 02138, U.S.A.
         \and
                Department of Astrophysics, Radboud University,
                Toernooiveld 1, 6525 ED Nijmegen, the Netherlands
         \and
                Department of Astronomy, University of Maryland,
                College Park, MD 20742, U.S.A.
         \and
                Astroparticle Physics Laboratory, Mail Code 661, NASA
                Goddard Space Flight Center, Greenbelt, MD 20771, U.S.A.
         \and
                MIT Kavli Institute for Astrophysics and Space Research,
                Cambridge, MA 02139, U.S.A.
          }

\date{Accepted by A\&A on August 23, 2009}

\abstract{An X-ray monitoring campaign with Chandra and Swift confirms
  that \bron\ is accreting at the lowest rate among the known
  persistently accreting low-mass X-ray binaries.  A thermonuclear
  X-ray burst was detected with the all-sky monitor on RXTE. This is
  only the second such burst seen in \bron\ in more than 20 Ms of
  observations done over 19 years.  The low burst recurrence rate is
  in line with the low accretion rate.  The persistent nature and low
  accretion rate can be reconciled within accretion disk theory if the
  binary system is ultracompact. An unprecedentedly short orbital
  period of less than $\approx7$ minutes would be implied. An
  ultracompact nature, together with the properties of the type~I
  X-ray burst, suggests, in turn, that helium-rich material is
  accreted. Optical follow-up of the Chandra error region does not
  reveal an unambiguous counterpart.

}

\maketitle

\section{Introduction}
\label{intro}

\bron, also known as RX J1718.4$-$4029 (and hereafter as J1718) is an
exceptional low-mass X-ray binary (LMXB). The measured X-ray flux
history of the source yields an inferred mass accretion rate that is
the lowest among all persistently accreting LMXBs with a proven
neutron star primary, by an order of magnitude \citep{zan07}.

J1718 was first detected in the ROSAT all-sky survey in 1990
\citep{vog99} and re-observed in 1994. Optical follow up of the X-ray
position failed to identify a counterpart \citep{mot98}. In 2000, a
thermonuclear X-ray burst was reported from observations with the
BeppoSAX Wide Field Cameras \citep[WFCs;][]{kap00}, establishing that
the source is an LMXB with a neutron star accretor. The burst was
relatively long in duration compared to typical bursts (i.e., longer
than 3 min) and exhibited photospheric radius expansion, implying that
the Eddington limit was reached during the peak of the burst.  For a
given Eddington luminosity and measured flux, one infers a source
distance of 6.5 kpc or 9 kpc for a hydrogen-rich or poor photosphere,
respectively \citep{kap00,zan05}.

A 15 ksec Chandra ACIS-S exposure taken in 2004 shows a 0.5--10 keV
flux of 5$\times10^{-12}$~\ecs\ corresponding to a luminosity of
$(5-9)\times10^{34}$~\lum\ after correction for interstellar
absorption \citep{zan05}. This is 0.03\% of the Eddington limit, which
is relatively low within the LMXB population. The latter fraction is
independent of distance or Eddington limit value, because it is simply
the flux normalized to the bolometric unabsorbed burst peak flux). The
low flux is roughly two times fainter than during the ROSAT
observations in 1990 and 1994, and is consistent with non-detections
by other instruments such as the BeppoSAX-WFC.  The X-ray absorption
is fairly high with an equivalent hydrogen number column density of
$N_{\rm H}=(1.3\pm0.1)\times10^{22}$~cm$^{-2}$ while the continuum
could be described by a power law with a photon index of
$2.1\pm0.2$. No modulation was detected in the Chandra-measured light
curve with an upper limit to the amplitude of 10\% (full bandpass) for
periods between 10 and 5000 s.

The low luminosity points to a low mass accretion rate. This is
supported by the fact that previously only a single burst was
detected. The burst recurrence time is longer than 15~d \citep{zan07},
which is at least two orders of magnitude longer than the recurrence
times measured for sources accreting at approximately 10\% of
Eddington \citep[e.g.,][]{gal08}. The implied mass accretion rate
($\la 0.1$\% of Eddington) is the lowest of all persistently accreting
LMXBs. We define 'persistent' here as accreting for longer than the
typical (few-months) viscous time scale of the accretion disk.  The
apparent persistent nature in combination with the low luminosity
suggests that J1718 is actually an ultracompact X-ray binary
\citep{zan07}. Ultracompact X-ray binaries (UCXBs) are characterized
by an orbital period shorter than $P_{\rm orb}\approx$80~min, implying
that the Roche lobe the donor star is so small that the donor must
have lost all or most of its hydrogen (Nelson et al. 1986; Savonije et
al. 1986). For the small accretion disk expected in such a binary it
is probable that even a low-luminosity X-ray source would keep the
disk entirely ionized and, therefore, the viscosity in the disk at a
level sufficient to sustain accretion onto the compact object. This
could allow the accretion disk instability to be avoided and the
source to emit persistently instead of transiently \citep{jvp96}.

No good-quality long-term light curve has been available for J1718,
precluding the accurate determination of the long-term average of the
flux. The source is too faint for current monitoring instruments and
programs. Therefore, we carried out a highly sensitive monitoring
program with Chandra and Swift. Furthermore, we performed optical
follow up of the previously determined position of the X-ray source
\citep{zan05} to search for an optical counterpart.  The results are
the subject of this paper.

\section{Observations}
\label{obs}

\begin{table}
\begin{center}
\caption[]{Itinerary of 4 Chandra and 48 Swift observations\label{tab1}}
\begin{tabular}{llrrl}
\hline\hline
ObsID & Date & Exposure & \#       & Off-axis \\
      &      & (sec)    & photons & angle (\arcmin) \\
\hline
\multicolumn{5}{c}{Chandra HRC-S} \\
\hline
6571 & 2006-05-16 & 5166 & 366 & 0.3 \\
6572 & 2007-01-28 & 5000 & 305 & 0.3 \\
\hline
\multicolumn{5}{c}{Chandra ACIS-S} \\
\hline
7464 & 2007-05-01 & 2423 & 537 & 0.3 \\
7465 & 2007-10-02 & 2517 & 613 & 0.3 \\
\hline
\multicolumn{5}{c}{Swift XRT} \\
\hline
35716001 & 2006-09-25 & 2493 &  136 & 2.181 \\
35716002 & 2007-01-28 & 3210 &  166 & 0.465 \\
90056001 & 2008-04-02 & 3320 &  541 & 0.852 \\
90056002 & 2008-04-09 & 2926 &  569 & 0.543 \\
90056003 & 2008-04-16 & 2061 &  261 & 0.754 \\
90056004 & 2008-04-23 & 1548 &  272 & 1.118 \\
90056005 & 2008-04-30 & 1603 &  186 & 1.183 \\
90056006 & 2008-05-07 & 2058 &  166 & 2.518 \\
90056007 & 2008-05-14 & 2267 &  104 & 1.411 \\
90056008 & 2008-05-26 & 1432 &   64 & 2.394 \\
90056009 & 2008-05-28 & 2137 &   54 & 1.257 \\
90056010 & 2008-06-04 &  678 &   19 & 2.438 \\
90056011 & 2008-06-07 &  466 &   17 & 3.968 \\
90056012 & 2008-06-11 & 1892 &   73 & 0.491 \\
90056013 & 2008-06-14 & 1420 &   47 & 2.180 \\
90056014 & 2008-06-21 & 2129 &   62 & 2.939 \\
90056015 & 2008-06-25 & 2380 &   79 & 1.904 \\
90056016 & 2008-06-28 & 1672 &   85 & 3.330 \\
90056017 & 2008-07-02 & 2055 &  104 & 2.995 \\
90056018 & 2008-07-05 & 2061 &  122 & 1.087 \\
90056019 & 2008-07-09 & 1339 &   46 & 2.069 \\
90056020 & 2008-07-12 & 1905 &   72 & 0.771 \\
90056021 & 2008-07-19 & 2371 &  107 & 2.488 \\
90056022 & 2008-07-23 & 2096 &   81 & 2.888 \\
90056023 & 2008-07-26 & 2273 &  102 & 2.339 \\
90056024 & 2008-07-30 & 1981 &   71 & 2.764 \\
90056025 & 2008-08-02 & 2441 &  126 & 1.996 \\
90056026 & 2008-08-06 & 1992 &   64 & 2.271 \\
90056027 & 2008-08-13 & 2286 &   98 & 1.846 \\
90056028 & 2008-08-20 & 2224 &   83 & 1.781 \\
90056029 & 2008-08-27 & 2159 &  113 & 2.305 \\
90056030 & 2008-09-03 & 3282 &  129 & 2.163 \\
90056031 & 2008-09-10 & 1816 &   69 & 3.620 \\
90056032 & 2008-09-17 & 2043 &   92 & 3.889 \\
90056033 & 2008-09-24 & 2183 &  103 & 2.884 \\
90056034 & 2008-10-01 & 1008 &   50 & 2.815 \\
90056035 & 2008-10-08 & 2245 &  104 & 1.122 \\
90056036 & 2008-10-15 &  816 &   17 & 3.246 \\
90056037 & 2008-10-21 & 1680 &   63 & 4.714 \\
90056038 & 2008-10-29 &  522 &   22 & 4.087 \\
90056039 & 2009-01-28 & 2208 &  145 & 2.072 \\
35716003 & 2009-01-30 & 3189 &   96 & 1.346 \\
90056040 & 2009-02-04 & 2125 &  136 & 1.870 \\
90056041 & 2009-02-11 & 2320 &   86 & 0.233 \\
90056042 & 2009-02-25 & 1633 &   90 & 1.963 \\
90056043 & 2009-03-04 & 1783 &  114 & 0.830 \\
90056044 & 2009-03-11 & 1845 &   64 & 1.271 \\
90056045 & 2009-03-25 & 2317 &  132 & 1.348 \\
\hline\hline
\end{tabular}
\end{center}
\end{table}

Chandra observations were obtained on four occasions during 2006 and
2007. Later, the source was observed with Swift once a week over the
course of a year (between April 2008 and April 2009), except for a
period of 3 months when observations were done twice a week. The
observation log is presented in Table~\ref{tab1}.

Of the 4 Chandra observations, two were done with ACIS-S \citep{gar03}
and two with HRC-S \citep{zom95,mur98}. ACIS-S was employed for
spectral resolution and HRC-S for time resolution.  The two Chandra
ACIS-S observations were carried out in the Timed Exposure mode with
ACIS CCDs S1-S4 on, with the aimpoint on the back-illuminated S3
chip. We selected a 128-rows sub-array starting at row 449. The
resulting time resolution is 0.741~s. The spectral resolution is
roughly 5\% (FWHM). The detector resolution is
$0\farcs49\times0\farcs49$ while the mirror point-spread function
(PSF) has a half-power density radius of 0\farcs5 for off-axis angles
smaller than 1\farcm5. The HRC-S multi-channel-plate detector was
chosen for two Chandra observations to obtain the best time resolution
($2^{-16}$ s$\approx$0.153~$\mu$s) at the expense of nearly all energy
resolution. The detector pixel size is $0\farcs13\times0\farcs13$ in a
$6\arcmin\times99\arcmin$ field of view.

The Swift-XRT observations were all carried out in Photon Counting
mode. This preserves the full imaging resolution
($2\farcs36\times2\farcs36$ over a $19\arcmin\times19\arcmin$ field of
view with a half-power diameter of the PSF of 18-22\arcsec) while the
time resolution is 2.5 s.  The total exposure time obtained with
Swift-BAT (15-150 keV) up to April 3, 2009, is 11.4
Msec \footnote{http://swift.gsfc.nasa.gov/docs/swift/results/transients/}.
Not a single detection was obtained in these observations, in daily or
orbital exposures.

J1718 is monitored by the All-Sky Monitor (ASM) on RXTE. This
instrument consists of three Shadow Scanning Cameras (SSCs), sensitive
between 1.5 and 12 keV, that dwell at a certain position on the sky
for 90~s after which they rotate slightly to a neighboring position
and dwell there for 90~s \citep{lev96}. Up to March 26, 2009, the ASM
carried out 37470 dwells with J1718 in the field of view of at least
one SSC. The total exposure time of these dwells amounts to about 3.4
Msec. The 3$\sigma$ detection threshold is about 300 mCrab per dwell
which translates to $6\times10^{-9}$~\ecs\ (2-10 keV). J1718 was
detected in two dwells, see Sect.~\ref{asmburst}. No other X-ray
<instrument was observing J1718 at the time of these dwells.

The total exposure time obtained with BeppoSAX-WFC (2-28 keV) is 7.20
Msec. These data reveal the detection of one burst \citep{kap00} while
any non-burst emission is below the detection threshold.

There is no mention of any detection of J1718 with INTEGRAL, for
instance in the most recent IBIS catalog \citep{bir07} nor in
data that are available
online\footnote{http://isdc.unige.ch/index.cgi?Data+sources,
  http://isdc.unige.ch/Science/BULGE/}.  We estimate that the source
was in the ISGRI field of view for 5870 science windows which would
translate to approximately 12 Msec up to October 24, 2006.

J1718 is covered by the semi-weekly Galactic bulge scan program
performed with the RXTE-PCA \citep{swa01,mar06}. Between mid 2004 and
March 31, 2009, 347 scans were performed, yielding $\approx$10.4 ks
exposure time. The source was never detected above a detection
threshold of $3\times10^{-11}$~\ecs\ (2-10 keV).

J1718 was observed in the optical with the Inamori-Magellan Areal
Camera and Spectrograph (IMACS) on the 6.5\,m Magellan-Baade telescope
on July 7, 2005, under photometric conditions with $0\farcs75$
seeing. Of the eight 4k$\times$2k detectors we only analyzed the
detector containing the X-ray source position. Observations were
obtained with $2\times2$ binning, giving a pixel scale of
$0\farcs22$\,pix$^{-1}$. Two 5\,min images in the $I$-band were
obtained, as well as two shorter images of 2 and 10\,sec exposure
times. The standard field SA\,110 \citep{lan92} was observed for
photometric calibration. All images were corrected for bias and
flat-fielded using domeflats according to standard methods.

\begin{figure}[t]
\includegraphics[width=\columnwidth,angle=0]{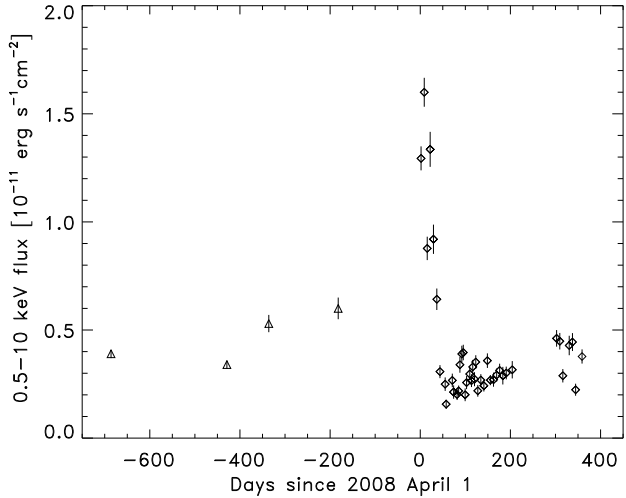}
\caption{Light curve of \bron\ (0.5-10 keV flux, uncorrected for
  absorption). The Chandra data points (triangles) were taken before
  April 1, 2008, and the Swift-XRT data points (diamonds) afterwards
  (see Table~\ref{tab1}). Vertical lines show 1$\sigma$ uncertainties
  \label{figlctime}}
\end{figure}

\section{Data analysis}
\label{ana}

\subsection{Chandra}

We employed {\tt ciao} version 4.1.1 for our analysis. For the ACIS-S
observations, source photons were extracted from a circular region
with a radius of $1\farcs8$ resulting in the numbers of photons listed
in Table~\ref{tab1}. For the background region photons were extracted
in an annulus centered on the source position with an inner radius of
$4\farcs9$ and an outer radius of $24\farcs5$. The spectra and
response files were extracted with the standard {\tt ciao} analysis
threads for a CCD temperature of $-$120 $^{\rm o}$C and the spectral
bins were grouped so that each bin contains at least 15 photons to
ensure applicability of the $\chi^2$ statistic.  The spectra between 1
and 8 keV are satisfactorily modeled by an absorbed power law, with
the hydrogen column density $N_{\rm H}$ fixed to the value
$1.3\times10^{22}$~cm$^{-2}$ found by \cite{zan05}. The power law
photon index is $2.37\pm0.11$ (1$\sigma$ error) for ObsID 7464 and
$2.44\pm0.10$ for ObsID 7465 ($\chi^2_\nu=1.17$ and 0.83,
respectively, with $\nu=30$ and 33). The unabsorbed 0.5-10 keV power
law flux is $(5.61\pm0.31)\times10^{-12}$ and
$(6.54\pm0.35)\times10^{-12}$~\ecs, respectively. We note that the
effect of pile up is smaller than the statistical uncertainty for a
CCD readout time of 0.741~s and a photon rate at the PSF peak of 0.05
s$^{-1}$.

For the HRC-S observations we employed a source extraction radius of
$2\farcs2$. There are no adequate response data available for the HRC
because of its extremely limited spectral resolution.  Therefore, we
refrain from a spectral analysis and merely use {\tt webpimms} to
calculate 0.5-10 keV fluxes based on the absorbed power-law spectrum
measured with ACIS-S.  The flux results from both HRC and ACIS-S are
given in Fig.~\ref{figlctime}. While the HRC data have fine time
resolution, we find no periodicity in the data. The upper limit on the
amplitude of a sinusoidal variation is approximately 20\% between 1
mHz and 1 KHz.

\subsection{Swift}

We exploited the 'cleaned' event files as supplied by the Swift
archive. Cleaning here refers to, for example, accounting for bad
pixels, coordinate transformations, reconstruction of events,
computation of energy channel values, elimination of partially exposed
pixels and filtering out times of bad attitude data. All events of
grades 0 to and including 12 were included in the analysis (the grade
refers to the configuration of pixel values per event).  After
examining each observation image, we found that a circular region of
radius 19\farcs8 encompasses most source photons. Formally, this size
is equal to that containing the half-power density of the
PSF. Exposure maps were constructed with {\tt xrtexpomap}, to allow
for a correction for bad CCD pixels.  The effective area was
calculated with {\tt xrtmkarf}.

For the Swift-XRT PSF and the readout time of 2.5 s in PC mode,
pile-up losses exceed the 5\% level for point source photon counting
rates of 0.32 s$^{-1}$ \citep{vau06}. The peak rate observed for J1718
is 0.2 s$^{-1}$ implying a pile-up loss of 3 to 4\%. This is similar
to the statistical uncertainty. For 45 out of 48 observations the
photon rate is less than 0.06 s$^{-1}$ implying pile-up losses of less
than 1\%, compared to statistical errors of about 10\%. A rough limit
for pile-up effects to become important in spectral analyses of XRT
data in PC mode is 0.6~c~s$^{-1}$ \citep{god09}. Thus, we neglect pile
up.

We extracted spectra for each observation and constructed response
matrices. Subsequently, we satisfactorily modeled the data with an
absorbed power law function. First, we modeled the high flux data and
low-flux data separately (ObsIDs 90056001-6 and the remainder
respectively), tying $N_{\rm H}$ to a single value for both
spectra. We find $N_{\rm H}=(1.23\pm0.08)\times10^{22}$~cm$^{-2}$, a
photon index of $\Gamma=2.18\pm0.11$ and $2.53\pm0.11$ for the high
and low-flux state, respectively with $\chi^2_\nu=1.135$ for
$\nu=190$. $N_{\rm H}$ is consistent with the value found by
\cite{zan05} ($1.32^{+0.16}_{-0.12}\times10^{22}$~cm$^{-2}$).  Fitting
a single value of $\Gamma$ to both spectra jointly results in
$\chi^2_\nu=1.274$ for $\nu=191$. The F-test yields a chance
probability to obtain such a change in $\chi^2$ by chance of
$2\times10^{-6}$, indicating that the photon index indeed changes
between the high and low state. Next, we modeled each observation
separately. For the low-state data we allowed only the power law
normalization to float freely. For the high state data the photon
index was also allowed to float freely. The resulting 0.5-10 keV
absorbed fluxes are shown in Fig.~\ref{figlctime}. These translate to
an unabsorbed 0.5-10 keV flux range of 0.4 to
3.5$\times10^{-11}$~\ecs.

\subsection{RXTE-ASM}
\label{asmburst}

\begin{figure}[t]
\includegraphics[width=\columnwidth,angle=0]{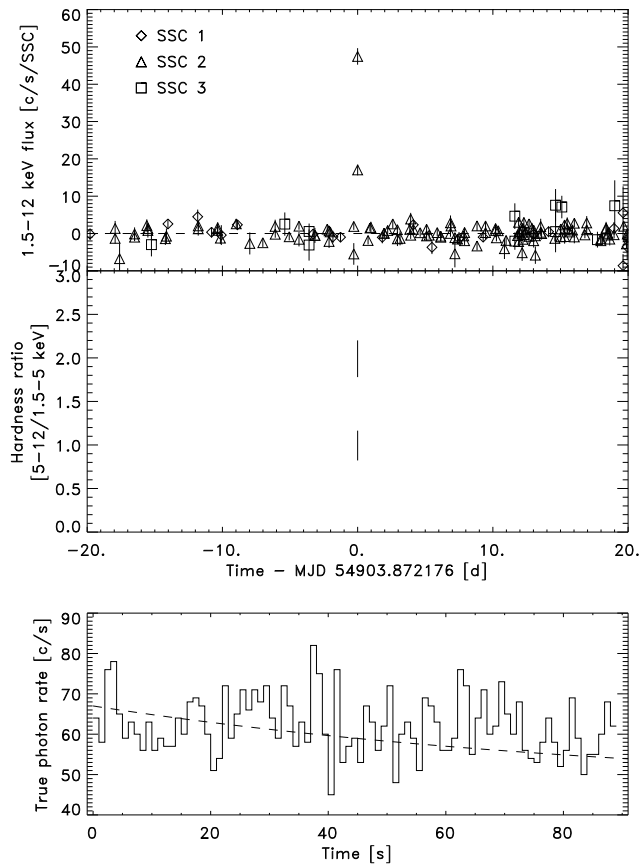}
\caption{{\em (top panel)} The photon flux for a standard SSC in March
  1996 \cite[this is the intensity renormalized to the whole detector,
    see][]{lev96}. The highest flux point occurs before the
  second-highest flux point. {\em (middle panel)} Spectral hardness
  defined as the ratio of the photon flux in the 5-10 keV band to that
  in the 1.5-5 keV band. The highest measurement corresponds to the
  highest flux point in the top panel. {\em (bottom panel)} The total
  true photon count rate of SSC2 at 1~s time resolution during the
  first dwell. The count rate includes contribution not only from
  J1718 (standard SSC flux 49 c~s$^{-1}$), but also from GX 340+0 (23
  c~s$^{-1}$), 4U 1702-429 (4 c~s$^{-1}$), 4U 1705-440 (7 c~s$^{-1}$),
  4U 1735-444 (10 c~s$^{-1}$), and background. The dashed curve shows
  an exponential function which is consistent with the dwell-to-dwell
  decay (i.e., with an e-folding decay time of 93 s and a flux that is
  corrected for the combined mask transmission and vignetting factor
  for this dwell of 0.298). \label{figburst}}
\end{figure}

A burst was detected with the RXTE-ASM on March 13, 2009, at 20:55 UT
(MJD~54903.872).  Figure~\ref{figburst} shows the photon flux
measurements of J1718 per ASM dwell in the full bandpass (1.5-12
keV). The source is bright in two consecutive dwells of SSC2 (SSCs
being numbered 1 through 3). Each dwell encompasses an exposure of 89
s and the gap between the two dwells is 7 s. J1718 was not detected in
the closest prior dwell to the burst, 6.7~h earlier, nor in the dwell
which follows the burst, 18.5~h later. We note that there was a PCA
slew over J1718 as part of the Galactic Bulge scan program (see
Sect. \ref{obs}) 1.68~h before the burst. The source was not detected
above a detection threshold of $3\times10^{-11}$~\ecs\ (2-10 keV).

The middle panel of Fig.~\ref{figburst} shows the spectral hardness
ratio defined as the ratio of the photon flux in the 5-12 keV band to
that in the 1.5-5 keV band (formally, the ratio of ASM channel 3 to
channels 1+2). If we model the hardness ratio by absorbed black body
radiation \citep[for a detailed example of a spectral ASM analysis,
  see][]{kee08} with $N_{\rm H}=1.3\times10^{22}$~cm$^{-2}$ (see
above), then the two measurements imply temperatures of k$T=3.3\pm0.3$
and $2.0\pm0.3$~keV. This cooling is what is expected from a
thermonuclear X-ray burst.

If one assumes that the decay from the first to the second dwell is
exponential, the implied e-folding decay time in the full bandpass is
$93\pm8$~s. Extrapolating the exponential decay backwards, a flux
level equal to that of the peak of the WFC-detected burst
\citep{kap00} is reached $23\pm4$~s prior to the start of the first
high dwell (the photon flux in the peak ASM dwell is a factor of 1.9
lower than the peak WFC flux). Assuming a similar behavior, the burst
probably started between that time and one minute earlier since the
WFC burst started with a 1-min long flat phase at the peak flux.

The total count rate detected in SSC2 during the first dwell (bottom
panel of Fig.~\ref{figburst}) does not show a clear decay. Comparing
the total rates in the two halves of this dwell, they differ by an
insignificant 2.3$\sigma$. The exponential decay expected from the
burst in this first dwell (dashed curve) is not clearly detected, but
it is consistent with the data ($\chi^2_\nu=1.005$ for $\nu=87$).

Given the peculiarity of the source and the lack of an observation
simultaneous to the first X-ray burst sensitive enough to detect the
persistent emission, one might question whether the ROSAT source is
really the origin of the X-ray burst. The burst might have occurred
from a nearby transient source. However, we argue against that
interpretation for the following reasons. First, no other bright
source was ever detected within the 0\farcm9-radius error region of
the first burst, even including the highest spatial resolution and
highly sensitive Chandra observations (the deepest detection threshold
is provided by the 2004 Chandra observation and has a value of
$3\times10^{-14}$~\ecs\ in 0.5-10 keV). Second, there is a sensitive
imaging measurement only 2.15 d before the second burst with Swift
(ObsID 90056044) and it does not show any other bright source within
10\farcm 0. We believe these two findings enforce the association
between the bursts and J1718. The thermonuclear nature of the first
burst identifies the accretor as a neutron star and the binary as a
LMXB. One possible alternative explanation, an X-ray flash
\citep{hei06}, is much less likely because X-ray flashes do not
exhibit thermal spectra, nor do they repeat.

\subsection{Optical}
\label{opt}

\begin{figure}[t]
\includegraphics[width=\columnwidth,angle=0]{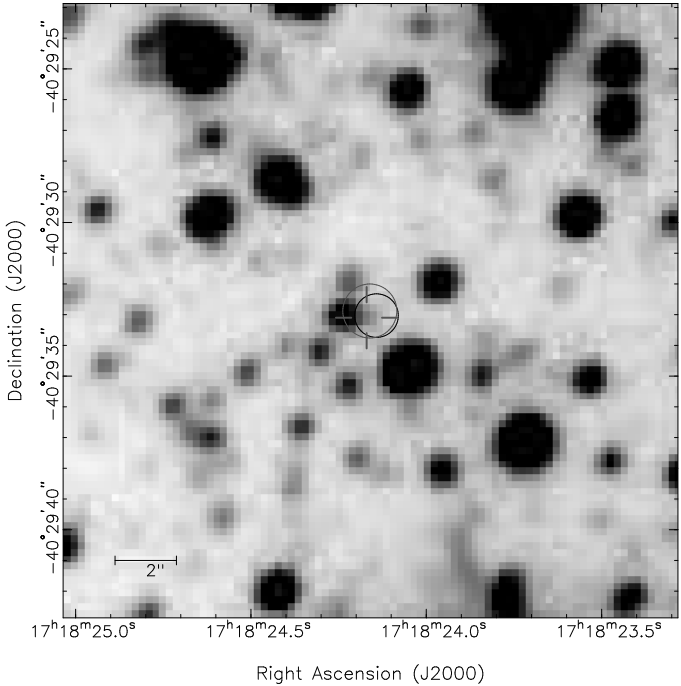}
\caption{IMACS image in the $I$-band taken on July 7, 2005, with a
  seeing of 0\farcs75. The circles are the 95\%-confidence error
  regions for the X-ray source with and without boresight correction
  (large and small circle, respectively). The open cross is centered
  on the position of the candidate optical
  counterpart.\label{figoptical}}
\end{figure}

Astrometric standards from the 2nd version of the USNO CCD Astrograph
Catalog \citep{zuz+04} were matched against stars on the 2\,sec
$I$-band image. Of the 21 stars overlapping with the image, three are
saturated. The remaining stars yield an astrometric solution with rms
residuals of $0\farcs059$ in right ascension and $0\farcs035$ in
declination. This calibration was transferred to the average of the
two 5\,min exposures using some 650 secondary standards determined
from the 2\,sec image. This gives rms residuals of $0\farcs019$ in
both right ascension and declination.

From the positions of X-ray sources detected with the ACIS S3 chip in
2004 by the \texttt{wavdetect} algorithm \citep{zan05}, we find that
the 5th brightest X-ray source (14 counts) coincides with a bright
$V=11.2$ star (2UCAC\,14672492). Based on the UCAC2 position of this
star, the boresight of the X-ray frame needs to be corrected by
$\Delta \alpha=0\farcs21\pm0\farcs30$ and $\Delta
\delta=0\farcs15\pm0\farcs18$. Though the boresight correction is not
significant by itself, we show in Fig.\,\ref{figoptical} both the
corrected error circle and the uncorrected error circle. The
boresight-corrected error circle has a 95\% confidence radius of
$0\farcs88$. The uncorrected error circle, assuming a $0\farcs6$
uncertainty in the boresight (90\% confidence, \citealt{akc+00}), has
a radius of $0\farcs71$ (95\% confidence). Both error circles contain
a single star located at
$\alpha_\mathrm{J2000}=17^\mathrm{h}18^\mathrm{m}24\fs170\pm0\farcs06$
and $\delta_\mathrm{J2000}=-40\degr29\arcmin33\farcs10\pm0\farcs04$.
The probability that a physically unrelated star could appear by
chance in the 0\farcs71-radius Chandra error circle is 16\%, which is
relatively high).

Instrumental magnitudes were determined from the 10\,sec image and the
average of the two 5\,min images through PSF fitting using DAOphot II
\citep{ste87}. These instrumental magnitudes were calibrated against
19 photometric standards from the SA\,110 standard field, using
calibrated magnitudes from \citet{ste00}. Only a zero-point was
fitted, since no color information is provided by the present single
filter observations. The rms residual on the zeropoint was 0.1\,mag.
In the absence of extinction coefficients, no correction for the
difference in airmass was made. We note, though, that the difference
in airmass between the standard and science field is only
0.12\,airmass, and that a typical $I$-band extinction coefficient of
0.1\,mag\,airmass$^{-1}$ would lead to a change in the zeropoint of
only 0.01\,mag. An aperture correction was determined from isolated
bright stars in the 10\,s $I$-band image, and the photometric
calibration was then transferred to the average of the two 5\,min
images. Finally, we compared our calibrated $I$-band magnitudes with
Gunn $i$-band magnitudes of some 700 stars from the
DENIS\footnote{http://cdsweb.u-strasbg.fr/denis.html} catalog, and
found an offset in magnitude of only $0.09\pm0.15$\,mag. We note that
the $3\sigma$ detection threshold of the image is $I_{\rm thr}=23.8$
at unconfused locations.

The star located inside both error circles has $I=21.9\pm0.2$, taking
into account the $0.1$\,mag uncertainty in the zeropoint. For the
hydrogen column density of $N_H=1.3\times10^{22}$\,cm$^{-2}$, the
model by \citet{ps95} predicts $A_V=7.3$ or $A_I=4.3$ (using the
extinction coefficients by \citealt{sfd98}). For the distance of
9\,kpc, the absolute magnitude of the star is
$M_I=21.9-4.3-14.77=2.8$.

\section{Discussion}

We find that J1718 appears to be persistently active for at least 19
years. The unabsorbed flux level in the 0.5 to 10 keV band was never
measured to be above $3.5\times10^{-11}$~\ecs. The average unabsorbed
flux is $8.6\times10^{-12}$~\ecs which translates to an average
luminosity of $8\times10^{34}$~(d/9~kpc)$^2$~\lum. Assuming a
bolometric correction of a factor of 2 \citep[e.g.,][]{zan07}, the
ratio of the luminosity to the Eddington limit is
$\gamma=0.0004$. This confirms the unique status of this
object. Assuming a radiation efficiency between 10 and 100\%, the
accretion rate is between 0.04 and 0.4\% of the Eddington limit. J1718
is the persistent LMXB with the lowest mass accretion rate currently
known.

The detection of a second burst from J1718 provides a handle on the
recurrence time and accretion rate. If we combine the exposure times
of all most relevant instruments that have been or are sensitive to
X-ray bursts from J1718 (7.2 Msec with BeppoSAX/WFC, 3.4 Msec with
RXTE/ASM, 1 Msec with INTEGRAL/JEM-X, and 10 Msec with HETE-II/WXM),
the average wait time between bursts is $t_{\rm wait}=125\pm88$~d.
For a pure helium burst, $\alpha=120$, where $\alpha$ is the ratio of
the gravitational energy released per nucleon (195 MeV for a canonical
NS) to the nuclear energy released \citep[1.6 MeV per nucleon for
  helium burning; see review by][]{lew93}. For a burst energy of
$E_{\rm b}=(2.8\pm0.5)\times10^{40}$~erg, the rate of the released
gravitational energy is $\alpha\;E_{\rm b}/t_{\rm
  wait}=(3.1\pm2.3)\times10^{35}$~\lum. This is consistent with the
measured average bolometric luminosity we report above. The
measurements translate to a radiation efficiency between 30\% and
100\%, or an inclination angle $i$ between 0 and 73$^{\rm o}$ (if the
apparent luminosity decreases by a projection factor perpendicular to
the line of sight of cos$i$).

From the persistent nature and low luminosity it is possible to obtain
a constraint on $P_{\rm orb}$ using accretion disk
theory. \cite{jvp96} defines a threshold for the accretion luminosity
as a function of $P_{\rm orb}$ above which the accretion process is
likely to be persistent (see also Sect.~\ref{intro}): log($L_X{\rm
  )}=35.2+1.07\;{\rm log(}P_{\rm orb}/{\rm 1~hr)}$ (for neutron star
accretors). According to this relationship, J1718 should have a period
$P_{\rm orb}\la1$~hr. However, that relationship has been refined by
\cite{las08} and previous to that by \cite{del03} to log($L_X{\rm
  )}=36.1+1.61\;{\rm log(}P_{\rm orb}/{\rm 1~hr)}$ for a
solar-composition disk and log($L_X{\rm )}=37.4+1.67\;{\rm log(}P_{\rm
  orb}/{\rm 1~hr)}$ for a pure helium disk. Lasota et al. express
these relationships in terms of mass accretion rate which we translate
to luminosity assuming a gravitational energy release per nucleon of
195 MeV, a 100\% radiation efficiency and a mass ratio between the
donor and accretor $q<<1$. The threshold luminosities in these refined
expressions are up to 2 orders of magnitude higher for $P_{\rm
  orb}<1$~hr. The luminosity derived from the burst energetics of
$(3.1\pm2.3)\times10^{35}$~\lum\ is consistent with a persistently
accreting helium disk only if $P_{\rm orb}$ is smaller than 6 min
(1$\sigma$ confidence), or 7 min (2$\sigma$ confidence). For a
carbon-oxygen-rich disk, this limit would be even more extreme because
of higher ionization potentials of these atoms
\citep{las08,men02}. For a hydrogen disk $P_{\rm orb}$ would be
smaller than $35$~min, which is impossible because a hydrogen dwarf
would not fit within such an orbit \citep[e.g.,][]{nrj86,nel08}. The
luminosity derived from the average flux yields an even more extreme
limit. The 6-7 min upper limit is shorter than the 11~min for the
present low record holder 4U 1820-30 \citep{stella87}. However, a word
of caution is appropriate. \cite{las08} point out that three UCXBs do
not adhere to the above thresholds (4U 1850-087, M15 X-2 and 4U
1916-05) and consider small amounts of hydrogen in the donor
atmosphere as the most likely explanation. That explanation meets
another challenge though: the $\approx$20 min orbital periods for 4U
1850-087 \citep{hcn+96} and M15 X-2 \citep{dkz+05} preclude the
presence of any hydrogen \citep{nrj86}.

If the mass transfer rate is governed by angular momentum losses due
to gravitation radiation, the low rate of
$\approx2\times10^{-11}$~M$_\odot$~yr$^{-1}$ suggests a donor mass
less than 0.03~M$_\odot$ for an orbital period of less than 1 hr and a
NS mass of 1.4 M$_{\odot}$, which points to a hydrogen-poor white
dwarf donor star (\citealt{ver93}). If the burst is fueled by helium,
the donor probably is a white dwarf with a helium-rich atmosphere
which would imply that the binary came into contact after the donor
left the main sequence \citep[e.g.,][]{zan05}.

Measurement of the brightness of an optical counterpart can aid the
confirmation or rejection of an ultracompact binary
hypothesis. \cite{jvp94} determine the relation between the absolute
visual magnitude $M_V$, the accretion rate in terms of the Eddington
limit $\gamma$ and the orbital period $P_{\rm orb}$ for actively
accreting LMXBs. They find log($P_{\rm orb}/1~{\rm
  hr)}=-(0.661\pm0.101)\;M_V-0.75\;{\rm
  log}\gamma+(1.037\pm0.216)$. The calibration of this relation
includes UCXBs. The contribution of donor stars in UCXBs to the
optical flux is expected to be negligible since white dwarfs are
thought to have absolute visual magnitudes fainter than about 8
\citep[e.g.,][]{ber95}. Assuming a blue B0-A0 stellar spectrum for the
active accretion disk \citep{jvp95}, $M_V=M_I$ and
$\gamma=0.0008\pm0.0006$ translate to log($P_{\rm orb}/1~{\rm
  hr)}=1.51^{+0.45}_{-0.19}$. This is inconsistent with the source
being a UCXB. Therefore, and because of the high probability (16\%) of
chance alignment of an optical source with the X-ray source, the case
for this star being the optical counterpart is not strong. The
unconfused detection threshold is, for the same $N_{\rm H}$ and
distance as J1718, equivalent to $M_I=4.7$. At the location of J1718
the detection threshold is higher because there is a bright star
immediately next to the error region. For $\gamma=0.0008\pm0.0006$ and
$P_{\rm orb}<1$~hr, one expects $M_V>5.3\pm0.5$. For the assumed
spectral shape for B0-A0, $M_I\approx M_V>5.3$. This is below the
detection threshold. Therefore, better seeing and deeper imaging are
likely to be required to detect the optical counterpart. One would at
least have to obtain images with limiting magnitudes of $I>25$ to
search for the counterpart. A spectrum of the current candidate
counterpart could be useful to verify that it is not the spectrum of
an accretion disk. For instance, if the star is a cool main sequence
star, the Ca II triplet near 8500\AA\ is by far the strongest feature
in the red/near-IR spectrum \citep{mun02}.

J1718 is associated with the class of 'Burst-Only Sources'
\citep{coc01,cor02,zan05} or 'Very Faint X-ray Transients'
\citep['VFXT';][] {mun05,wij06,deg09}. \cite{kin06} suggest that these
transients either have hydrogen-poor donors, were born with very low
companion masses, or evolve more quickly than standard theory
predicts. A VFXT particularly similar to J1718 is XMMU
J174716.1-281048 \citep[e.g.,][]{sid03,san07b,san07a}: since 2003, it
has been seen at the faint level of
$\approx3\times10^{-12}$~\ecs\ (2-10 keV). Before that it was
quiescent as shown by several XMM-Newton and Chandra observations).  A
long X-ray burst was detected \citep{bra06} and the persistent
spectrum is a power law with a photon index of 2.1
\citep{san07a}. There may be more (quasi) persistent X-ray sources
that are similar in nature to J1718 but have not exhibited an X-ray
burst yet (remember that only two bursts were seen in J1718 in the
course of 19 years despite more than 20 Msec of monitoring
observations, see \S~\ref{obs}). Possible examples are 4U 1543-624 and
4U 1556-605 \citep{far03}. The similarity to VFXTs and other
unidentified persistent sources illustrates that sources like J1718
may be rather common. This has not been fully appreciated yet, because
it is difficult to catch the rare and brief X-ray bursts. Operation of
surveying X-ray programs and all-sky X-ray monitors is crucial in
advancing our understanding of these objects.

\section{Conclusion}

We confirm the LMXB nature of J1718 through the detection of a second
X-ray burst in the absence of any other bright source in the
neighborhood. In addition, we have determined an accurate value for
the {\em long-term average} flux of J1718. This value translates to
the lowest luminosity known for a {\em persistently} accreting
LMXB. This luminosity is consistent with accretion disk theory if the
orbital period is less than about 7 min. We find a candidate optical
counterpart that is too bright for a UCXB. A spectral study of this
counterpart is needed to check whether it is consistent with an UCXB,
or deep optical images, reaching at least $I=25$, obtained under
seeing conditions better than $\approx$0\farcs75, to search for a
weaker nearby counterpart.

\acknowledgement

We thank the anonymous referee for making excellent suggestions for
improving the paper.

\bibliographystyle{aaa} \bibliography{aa12403}

\end{document}